\documentclass[12pt,english]{article}
\usepackage[T1]{fontenc}
\usepackage[latin2]{inputenc}
\usepackage{geometry}
\usepackage{babel}
\usepackage{graphics}
\usepackage{floatflt}

\begin{document}

\title{Multiperiodic RR Lyrae stars in the center of the Galaxy}

\author{\textbf{Tomasz Mizerski}}

\maketitle

{\par\centering Warsaw University Observatory, Aleje Ujazdowskie 4, 00-478 Warszawa, Poland\par}

\vspace{0.5cm}

{\par\centering e-mail: mizerski@astrouw.edu.pl\par}

\vspace{0.5cm}

\begin{abstract}

This paper presents results of a fully automated search for multiperiodic RR Lyrae stars in
the center of the Galaxy.
The search was carried out on OGLE-II database created by means of image subtraction method.
I found more than 2700 RR Lyrae stars with about 600 showing various
kinds of multiperiodic behavior. Previous OGLE-I database contained only
about 200 RR Lyrae stars in the Galactic Bulge, that is twelve times less
than the current sample. There are two most interesting outcomes for Bulge: 
very high percentage of Blazhko stars and very low percentage of RRd stars.
Blazhko effect was confirmed in about 25 percent of all RRab stars. This incidence rate is almost 
2.5 times greater than in the Large Magellanic Cloud. On the other hand I discovered only 
3 RRd stars, which implies that the incidence rate of this type of variables is much smaller 
in Bulge than in the Magellanic Clouds.\\

\textbf{Key words}: pulsating stars: RR Lyr stars, stellar pulsation: Blazhko effect, radial and nonradial modes.
\end{abstract}

\section{Introduction}

\par \hspace{1cm} It is well known that RR Lyrae stars play very important role in modern
astrophysics. They are used as distance indicators, their spatial
distribution can provide us with useful information about the structure of
the Galaxy and they help us understand stellar evolution and
pulsation. Especially pulsation theory benefits from investigation of
large number of RR Lyrae stars after the discovery of nonradial
oscillations in these stars. Bulge catalog presented in this paper contains over 700 RR Lyrae 
stars showing multiperiodic behavior. In many cases such behavior can only be explained 
by excitation of nonradial pulsation modes. 

\section{Observational data}

\par \hspace{1cm} In the years 1997 - 1999, during the second phase of the Optical
Gravitational Lensing Experiment - OGLE-II, 49 fields in the center of the
Galaxy were monitored. The fields covered approximately 11 squared degrees
with most of them being located in regions of high stellar density. 
With the aid of Warsaw 1.3-m telescope at Las Campanas Observatory
of the Carnegie Institution of Washington about 200 observations in the \emph{I} band
were made for each field, mostly in driftscan mode. 
Technical details of the telescope are given in Udalski, Kubiak \& Szymański (1997).
The photometry was obtained using Difference Image Analysis software written by Woźniak
(2000).

\section{Data analysis}

\par \hspace{1cm} Image subtraction method preclassified as variable stars about 4500 objects
per field. Since there are 49 fields in OGLE-II database, selection of RR
Lyrae stars had to be automated. There are approximately 200 measurements
from three observational seasons for each star and the time span is about
1000 days. Observations are sampled unevenly. As the first step Lomb-Scargle periodograms, with frequency resolution (5\(\times\)\(10^{-4}\) c/d) for all stars were computed.
Such a resolution is high enough to classify a star as a periodic variable and to examine if it is a RR Lyr star.
At later stages periodograms for all selected RR Lyr stars were computed with much higher resolution.
After a periodogram was calculated for every star, the most probable period was selected out of 25 periods corresponding to the highest
peaks in its periodogram. For all of those 25 periods my algorithm fitted the data with a Fourier series with
various number of harmonic terms. This number ran from 1 to 6 for each
period which was done because of the diversity of RR Lyrae light curve shapes. Strongly
skewed curves of RRab stars require more harmonic terms than sinusoidal
curves of RRc stars. The criterion of selection was the smallest value of $\chi^{2}$ per degree
of freedom. The program always chose the pair {\em (period, number of harmonics)} that
minimized that quantity. As expected in the case of RRc stars
selected number of harmonic terms was mainly 1 or 2, but for RRab stars it
was typically 4 or 5 in most cases. 

\subsection{Identification of RR Lyrae stars}

\label{rrlsearch}

\par \hspace{1cm} When the most probable period and the best number
of harmonic terms were found for every star the algorithm shifted the zero point of every light curve in the
time domain to the maximum of the curve. The maximum was estimated from the Fourier fit. 
To separate RR Lyrae stars from all other variables the code computed the following parameters: 

\begin{equation}
A_{ij}=\frac{A_i}{A_j}
\end{equation}

\begin{equation}
\Phi_{ij}=j\cdot\phi_i-i\cdot\phi_j
\end{equation}

Quantities \(A_i\) and \(\phi_i \) are respectively the amplitude and the phase
of a cosine term with $\omega=i\cdot\omega_{0}$ where $\omega_{0}$
is the frequency of variability.
Parameters defined above, when combined
with the value of the period, allow not only to distinguish RR
Lyrae stars from other variables, but also to separate RRab and RRc classes. 
In the search for RRabs my code used parameters
\(A_{21}\), \(A_{32}\), \(A_{43}\) and \(\Phi_{21}\). 
For RRcs only \(A_{21}\), \(A_{32}\) and \(\Phi_{21}\) were used. 
Output RRab and RRc samples were contaminated by small amount of variables
of different types, most frequently High Amplitude $\delta$ Sct stars.
These were eliminated with the aid of color-magnitude diagrams and by inspection of 
all the light curves by eye. This method of selecting 
RR Lyrae stars was also used by Mizerski and Bejger (2001). 

\subsection{Detection of additional periodicities}

\par \hspace{1cm} After RR Lyrae stars were separated from all other variables 
by the algorithm described in subsection \ref{rrlsearch}, the search for multiperiodic behavior was performed. 
The whole process was fully automated and for each RR Lyrae star consisted of three main phases:
\begin{itemize}
\item Calculation of Lomb-Scargle periodogram with high resolution in frequency\\ (5\(\times\)\(10^{-6}\) c/d) and determination of
 the most probable period of variability.
\item Prewhitening of the light curve with the determined period with dynamically adjustable number of harmonic terms.
\item Calculation of Lomb-Scargle periodogram of the prewhitened light curve. 
\end{itemize}

In the first phase my code calculated Lomb-Scargle periodogram of the original light curve. This method 
of periodogram calculation works well for unevenly sampled data. Setting frequency resolution to 
5\(\times\)\(10^{-6}\) c/d allowed me to determine the first, main period
with high accuracy, considering that the time span of the observations was about 1000 days. Out of the periodogram 25 frequencies corresponding 
to the highest peaks were selected. For all those frequencies my code fitted the data by means of 
least squares method with a Fourier series. Number of harmonic terms was adjusted dynamically 
in a range between 1 and 6. The program searched for a pair (frequency,number of harmonics) which minimized 
the value of $\chi^{2}$ of the fit. After such a pair was found the inverse of the selected frequency was taken 
for the primary period of pulsation. The light curve was prewhitened with this model, which means that the Fourier 
series with frequency and number of harmonics that minimized $\chi^{2}$ was subtracted from the data. 
Then periodogram was calculated for the prewhitened data. 
If the periodogram of the prewhitened light curve showed presence of peaks 15 times greater than the median of 
the peaks the star was considered a multiperiodic candidate. The most probable value of the second period of 
variability was then determined in exactly the same way as in the case of the primary period. 
In a case of a star with two detected periods a two period model was fitted to the original light curve. The formula for such a fit
is given in equation \ref{dmod}. During this phase not only the amplitudes and phases were computed. Both periods 
were adjusted as well. The values of two periods of pulsation found in the earlier phases were the input values 
for the procedure. In the last step the model with two periods was
subtracted from the data and search for additional periodicities was performed. 

\par

\begin{equation}
F(t_k)=A_{00}+\Sigma_{j=1}^{m}A_{0j}\cos(j\omega_{2}t_{k}+\phi_{0j})+\Sigma_{i=1}^{m}
\Sigma_{j=i-m}^{m-i}A_{ij}\cos[|i\omega_{1}+j\omega_{2}|t_{k}+\phi_{ij}]
\label{dmod}
\end{equation}

\par

Number {\em m} is the rank of the fit. In a case where \emph{m}=2 the series
includes the constant term and the terms with frequencies \(\omega_1\), \(\omega_2\), 2\(\omega_1\),
2\(\omega_2\), \(\omega_1+\omega_2\), $\vert\omega_1-\omega_2\vert$.
Amplitude errors are computed as well, and only the terms
with amplitudes 3 times greater than their errors are included in the final fit.  
Value of \(\chi^2\) of a two period model strongly depends on the rank of
the fit. As an example Table \ref{tabdmod} demonstrates how $\chi^{2}$
values change with $m$ for a two period model of an RRd star. It this star
a model with two periods of variability is satisfactory, and fits the data
very well for $m \geq 4$. In the case shown in Table \ref{tabdmod} setting
$m>4$ does not reduce $\chi^{2}$, as the high order terms do not improve
the fit, while consuming additional degrees of freedom.

\begin{table}[h]

\centering{

\begin{tabular}[]{|c|c|c|c|c|c|c|}
\hline
m & 1 & 2 & 3 & 4 & 5 & 6 \\
\hline
$\chi^{2}$ & 30.154 & 5.342 & 2.249 & 1.505 & 1.661 & 1.641 \\
\hline
\end{tabular}

}

\caption{{\small \label{tabdmod}\emph{\(\chi^2\) values as a function of the rank of the fit.}}\small }

\end{table}

\section{Results. Statistics of multiperiodic variables.}

\par
\begin{figure}[h]
{\par\centering \resizebox*{7cm}{7cm}{\includegraphics{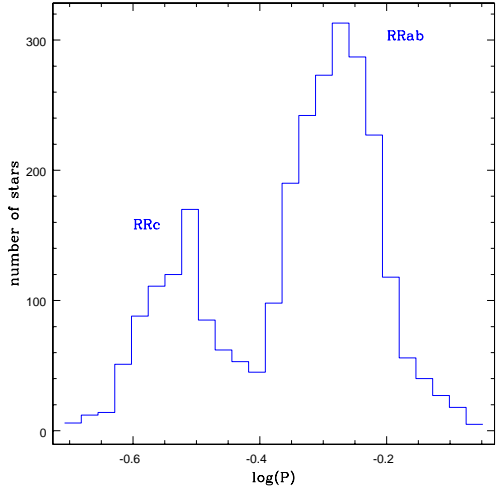}} \par}
\caption{{\small \label{hist}\emph{Period distribution of RR Lyrae stars in
the Galactic Bulge. Two populations are seen: RRab and RRc stars.}}\small }
\end{figure}

 \hspace{1cm} As the result I obtained an RRab sample containing 1942
stars and an RRc sample of 771 stars. This gives the total number of RR Lyraes equal to 2713,
which is much larger than 215 stars in OGLE-I RR Lyrae database.
Although my current sample is not as numerous as LMC sample of Alcock et
al. (1996) it
reveals a larger diversity of pulsation behaviors. Both RRab and RRc stars
cover a wide range of periods with $P_{ab}\in[0.340,0.922]d$ and
$P_{c}\in[0.201,0.446]d$. Average values of pulsation periods in both classes
are respectively $P_{ab}^{avr}=0.554d$ and $P_{c}^{avr}=0.308d$. Figure
\ref{hist} shows period histogram for all RR Lyrae stars. Presence of two
peaks reflects the fact of existence of two distinct populations of RR
Lyrae stars. These populations are fundamental mode pulsators RRabs and
first overtone pulsators RRcs.
 
Another interesting observation is a very high number of RR Lyrae stars that
exhibit multiperiodic behavior. I discovered about 800 such stars. Most of
them are Blazhko stars but there are also a few RRd stars, that is double mode
pulsators oscillating simultaneously in fundamental mode and first
overtone. Multiperiodic RR Lyrae stars will be discussed more thoroughly in the next subsections. 

\subsection{Double mode pulsators}

\par \hspace{1cm} RRd stars are very important in astrophysics because they provide us with one 
additional observable, that is the second period of variability. A widely known and intriguing fact concerning 
these variables is that they are very numerous in some stellar systems while almost completely absent in others.
This fact is not understood. So far there is no clear evidence that any
specific physical quantity is responsible for this difference. Moskalik
(2000) suggests that different incidences of RRd stars in stellar systems
could result from different metalicities in those systems. Many RRd variables were discovered in the Galactic disk and in some globular clusters 
in the Galaxy. Magellanic Clouds are also known to have quite numerous populations of RRd stars. Alcock et al. (2000) report 181 such 
variables in their LMC database. The fundamental question which arises from the analysis of OGLE data is why do we observe 
so very few RRd stars in the Galactic bulge. Previous investigations of
OGLE-I database performed by Moskalik (2000) and Pigulski et al. (2003) resulted in 
only two RRd variables found. Those were the only RRd stars known in this system until now. Analysis of OGLE-II database, 
which includes the fields of OGLE-I, led to the discovery of three additional RRd stars 
in the center of the Galaxy. Total number of only 5 known RRd variables in the Galactic bulge is in sharp contrast to the large 
number of objects of this kind observed in globular clusters and in the Magellanic Clouds. Incidence rate, defined as the ratio of
the number of RRd to the number of RRc variables, is $\approx 0.007$ for
the bulge whereas for the LMC Alcock et al. (2000)
give the value of $0.134$. This is obviously a striking difference: there are proportionally 20 times more RRd stars 
in the LMC than in the Galactic bulge. Unfortunately we still do not know why RRds are so rare in the center 
of the Galaxy.             

\par \hspace{1cm} Two of three RRd stars in OGLE-II database are very typical representatives of their class. The ratio of the periods of 
the first overtone and the fundamental mode, hereafter referred to as $\zeta$, is $\approx 0.745$ in both cases. The ratio of
the amplitudes of the second periodicity to the dominant periodicity, although substantially different for both stars, is in good 
agreement with the statistics given by Alcock et al. (2000). This ratio is denoted 
as $\xi$. Table \ref{tabrrd} gives the basic characteristics of the three
RRd stars I discovered, that is both periods $P_{0}$ and $P_{1}$, 
values of parameters $\zeta$ and $\xi$ and two values of $\chi^{2}$. The first value $\chi_{1}^2$ corresponds to a single period model
with only the dominant period present. The second value $\chi_{2}^2$ is the $\chi^2$ of a double period model with both periods of 
variability. These numbers are given for comparison to demonstrate that taking into account the second periodicity significantly reduces
$\chi^{2}$ of the fit. As it can be seen in table \ref{tabrrd}, the last
star, namely {\em bul\_sc39.1568}, differs from the first two
objects. It has a shorter period and a greater $\xi$ value than the two other
stars. It means that this star is separated from the other two on the
Petersen diagram, and therefore, according to Popielski et al. (2000), has
a different metallicity.   

\begin{table}[h]

\centering{

\begin{tabular}[]{|c|c|c|c|c|c|c|}
\hline
\emph{star} & $P_{0}$ & $P_{1}$ & $\zeta$ & $\xi$ & $\chi_{1}^2$ & $\chi_{2}^2$ \\
\hline
bul\_sc7.1529 & 0.487192 & 0.362842 & 0.744761 & 0.464 & 109.652 & 1.712 \\
bul\_sc21.7133 & 0.503068 & 0.374970 & 0.745366 & 0.176 & 55.805 & 3.384 \\
bul\_sc39.1568 & 0.461390 & 0.344906 & 0.747537 & 0.336 & 10.397 & 4.953 \\
\hline
\end{tabular}

          }

\caption{{\small \label{tabrrd} RRd stars.}}\small

\end{table}

\begin{figure}[h]
{\par\centering \resizebox*{16cm}{10cm}{\includegraphics{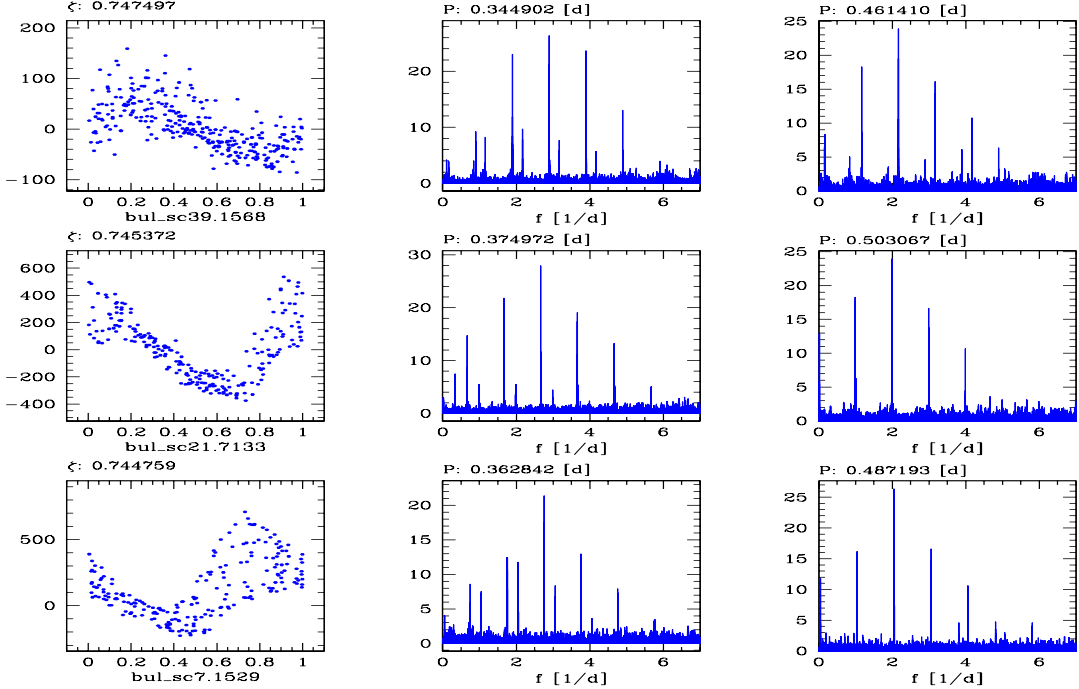}} \par}
\caption{{\small \label{diagrrd}\emph{RRd stars, placed in rows. The plots
show respectively: original light curve phased with the primary period
(first overtone), periodogram before and periodogram after prewhitening.}}\small }
\end{figure}

\subsection{Blazhko stars}

\par \hspace{1cm} Blazhko effect manifests itself in two different
ways. Prewhitened periodograms show either one or two peaks in the 
close vicinity of the primary period of pulsation. As for now it is
thought that these two behaviors divide Blazhko stars into 
two classes: BL1 class with frequency doublet and BL2 class with frequency
triplet. Members of BL1 class have their additional 
frequency very close to the frequency of the radial mode, typically closer 
than 0.01 [c/d]. In the case of BL2s the dominant frequency and the two
additional frequencies create an equidistant triplet centered on the
dominant frequency. This triplet is typically asymmetric, 
that is the heights of the additional peaks are different. 
So far we do not have a satisfactory and complete explanation of Blazhko 
effect. Most widely accepted theoretical approach explains the presence of 
additional peaks by excitation of nonradial modes. 
Those modes are very close in the frequency domain to the frequency of the radial mode.  
Existence of equidistant, almost symmetric frequency triplets could be
explained by other means but presence of strongly asymmetric 
triplets and especially doublets seem to require some kind of nonradial oscillations.

\par \hspace{1cm} It was thought for a long time that Blazhko effect exists
only among RRab stars. Recent studies indicate that it 
also concerns RRc variables. Moskalik \& Poretti (2002) report the discovery of two RRc
Blazhko stars. My research confirms that Blazhko effect is 
present in some RRc stars, although is not as frequent as in the RRab
class. Two next subsections cover consecutively RRab and RRc Blazhko 
stars found in the OGLE-II database. 

\subsubsection{RRab stars}

\par
\begin{figure}[h]
{\par\centering \resizebox*{16cm}{8cm}{\includegraphics{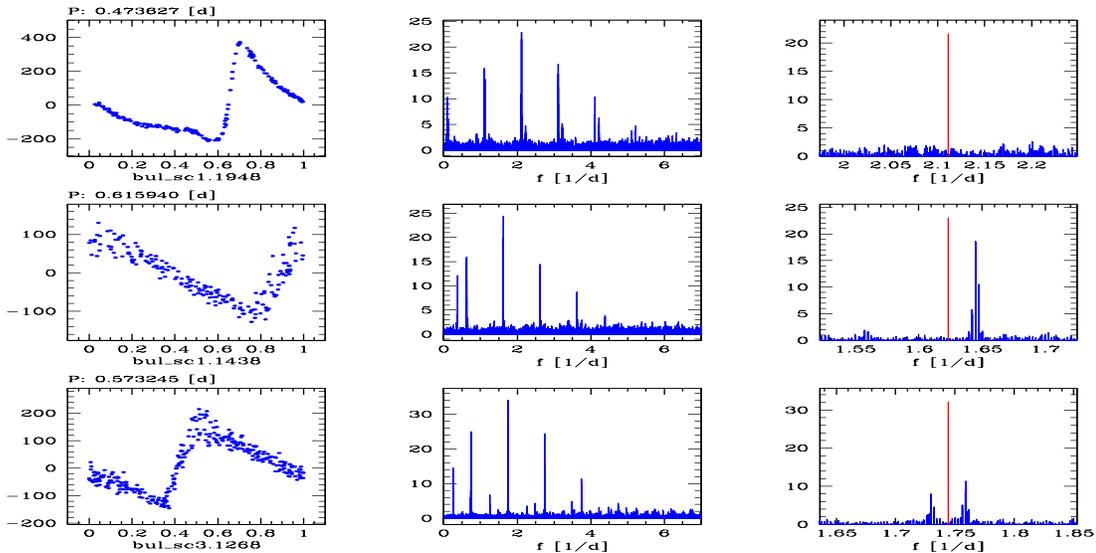}} \par}
\caption{{\small \label{bla}\emph{RRab Blazhko stars. Convention is the same as on figure \ref{diagrrd}. 
The line in the center of the prewhitened periodogram shows the location of the primary period. From the top:
star without additional peaks, BL1, BL2.}}\small }
\end{figure}

 \hspace{1cm} My code identified 243 BL1 and 143 BL2 type stars among
RRab stars.\\ BL1s constitute $12.5\%$ of all RRab in the OGLE-II database, while BL2s only $7.4\%$. Figure \ref{bla} shows two examples of such variables, one for
BL1 and one for BL2 classes.\\ A purely radial pulsator, with no additional
frequencies, was included in the figure for comparison.
Average frequency separation between the additional
peak and the radial mode is |$\Delta f=0.01$|. In the case of BL2s, that is stars with equidistant
triplets, $\Delta f$ is understood as the difference between the frequency
of the higher of the two subsidiary peaks and the frequency of the radial
mode. In the literature one can also find a quantity called modulation
period, which is the inverse of the frequency separation:
$P^{m}=\frac{1}{|\Delta f|}$.
\par \hspace{1cm} Basic properties of RRab Blazhko stars are as follows:

\begin{itemize}
\item BL1
\begin{itemize}
\item \(P_{min}\)=0.370d \(\;\) \(P_{max}\)=0.779d \(\;\) \(P_{a}\)=0.543d
\item \(P_{min}^{m}\)=5d \(\;\) \(P_{max}^{m}\)=968d \(\;\) \(P_{a}^{m}\)=91d
\item \(\Psi_{min}\)=0.039 \(\;\) \(\Psi_{max}\)=0.537 \(\;\) \(\Psi_{a}\)=0.160
\item \(\Delta\)f>0 in 81\(\%\) cases  
\end{itemize}
\item BL2
\begin{itemize}
\item \(P_{min}\)=0.362d \(\;\) \(P_{max}\)=0.680d \(\;\) \(P_{a}\)=0.521d
\item \(P_{min}^{m}\)=15d \(\;\) \(P_{max}^{m}\)=977d \(\;\) \(P_{a}^{m}\)=96d
\item \(\Psi_{min}\)=0.067 \(\;\) \(\Psi_{max}\)=0.595 \(\;\) \(\Psi_{a}\)=0.174
\item \(\psi_{min}\)=0.429 \(\;\) \(\psi_{max}\)=0.996 \(\;\) \(\psi_{a}\)=0.795
\item \(\Delta\)f>0 in 73\(\%\) cases
\end{itemize}
\end{itemize}

where $P$ represents radial pulsation period, $\Psi$ is the amplitude ratio of the additional peak to the peak
representing the radial mode, and $\psi$ is the amplitude ratio of the two
additional peaks in the case of BL2 stars. Subscripts {\em min, max, a}
denote minimal, maximal and average values respectively.
It can be seen that the amplitudes of additional peaks can be comparable
with the amplitude of the dominant peak, and that the triplets range from highly
symmetric to highly asymmetric. The fact that $\Delta f$ is positive
in most cases means that the additional peak in BL1s, or higher of the two
additional peaks in BL2s, corresponds to a higher frequency than the
frequency of the radial mode. This finding is in agreement with the results
of Moskalik \& Poretti (2002). As it will be shown in the next paragraph,
RRc Blazhko stars behave differently.

\begin{figure}[h]
{\par\centering \resizebox*{7cm}{7cm}{\includegraphics{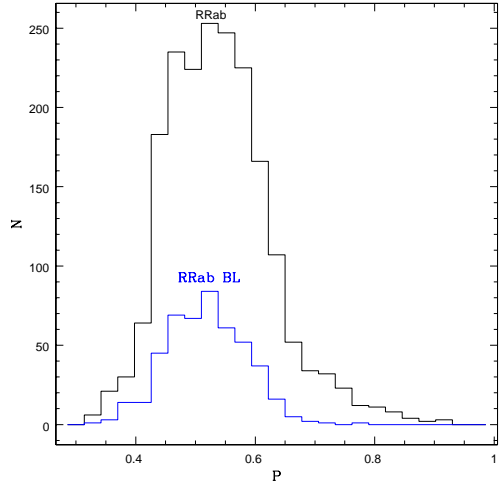}} \par}
\caption{{\small \label{bla_all}\emph{Period distributions for all RRab and RRab Blazhko stars.}}\small }
\end{figure}   

\hspace{1cm} Period distribution of RRab Blazhko stars is consistent with period
distribution of all RRab stars. Figure \ref{bla_all} demonstrates period
distributions of all RRab and RRab BL stars.

\subsubsection{RRc stars}

\par
\begin{figure}[h]
{\par\centering \resizebox*{16cm}{8cm}{\includegraphics{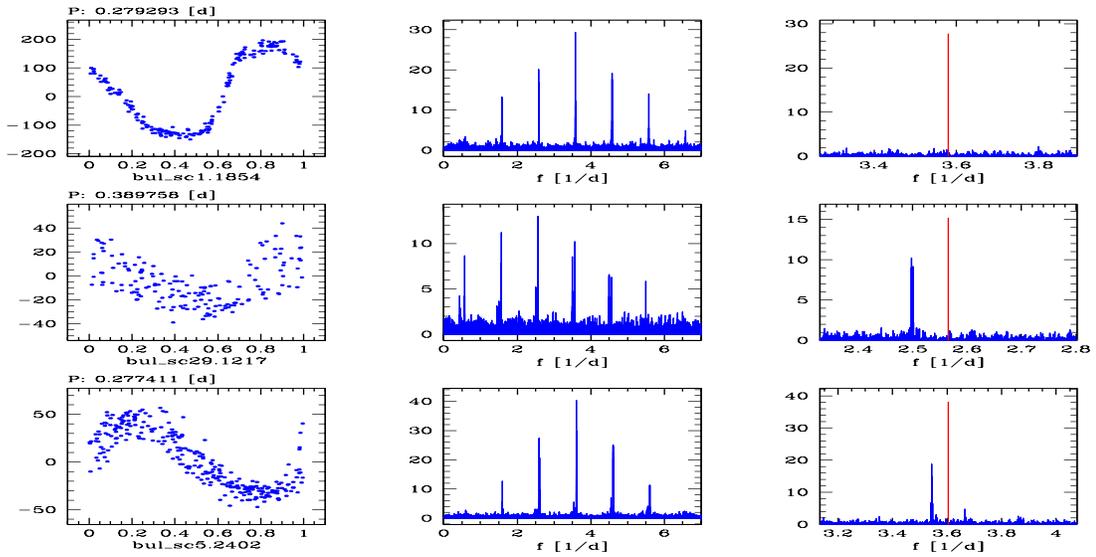}} \par}
\caption{{\small \label{blc}\emph{RRc Blazhko stars. Convention is the same as on figure \ref{bla}.}}\small }
\end{figure}

 \hspace{1cm} RRc Blazhko stars are less numerous than RRab Blazhko stars. I detected 22 BL1
and 30 BL2 RRc stars, and it is easy to see that the proportions are different
than those for RRabs. RRab Blazhko stars seem to prefer BL1 subclass, such
effect is not observable among RRc stars. The statistical properties of RRc
Blazhko stars differ from those of RRabs and are as follows:

\begin{itemize}
\item BL1
\begin{itemize}
\item \(P_{min}\)=0.246d \(\;\) \(P_{max}\)=0.390d \(\;\) \(P_{a}\)=0.288d
\item \(P_{min}^{m}\)=7d \(\;\) \(P_{max}^{m}\)=986d \(\;\) \(P_{a}^{m}\)=247d
\item \(\Psi_{min}\)=0.049 \(\;\) \(\Psi_{max}\)=0.820 \(\;\) \(\Psi_{a}\)=0.279
\item \(\Delta\)f>0 in 54\(\%\) cases 
\end{itemize}
\item BL2
\begin{itemize}
\item \(P_{min}\)=0.256d \(\;\) \(P_{max}\)=0.439d \(\;\) \(P_{a}\)=0.316d
\item \(P_{min}^{m}\)=2d \(\;\) \(P_{max}^{m}\)=913d \(\;\) \(P_{a}^{m}\)=397d
\item \(\Psi_{min}\)=0.054 \(\;\) \(\Psi_{max}\)=0.494 \(\;\) \(\Psi_{a}\)=0.178
\item \(\psi_{min}\)=0.426 \(\;\) \(\psi_{max}\)=0.999 \(\;\) \(\psi_{a}\)=0.844
\item \(\Delta\)f>0 in 33\(\%\) cases
\end{itemize}
\end{itemize}

with the symbols having exactly the same meaning as in the case of RRab stars.
As one can see there are substantial differences between RRab and RRc
Blazhko stars. Average modulation periods of RRc BLs are much longer and there is no
strong preference for $\Delta f>0$ among RRc BL1s. In fact for RRc BL2s we can observe an
opposite preference, $\Delta f$ is mostly negative, which means that the higher
of the additional peaks corresponds to a lower frequency than the radial
mode. Also the value of $\Psi_{max}=0.820$ for RRc BL1s seems to be an interesting
result, as it shows, that in the case of RRc Blazhko stars the nonradial mode can have
almost the same amplitude as the radial mode. 
Examples of RRc BL1s and BL2s are shown in figure \ref{blc}.

\subsection{Stars with multiple periods}

\par
\begin{figure}[h]
{\par\centering \resizebox*{16cm}{8cm}{\includegraphics{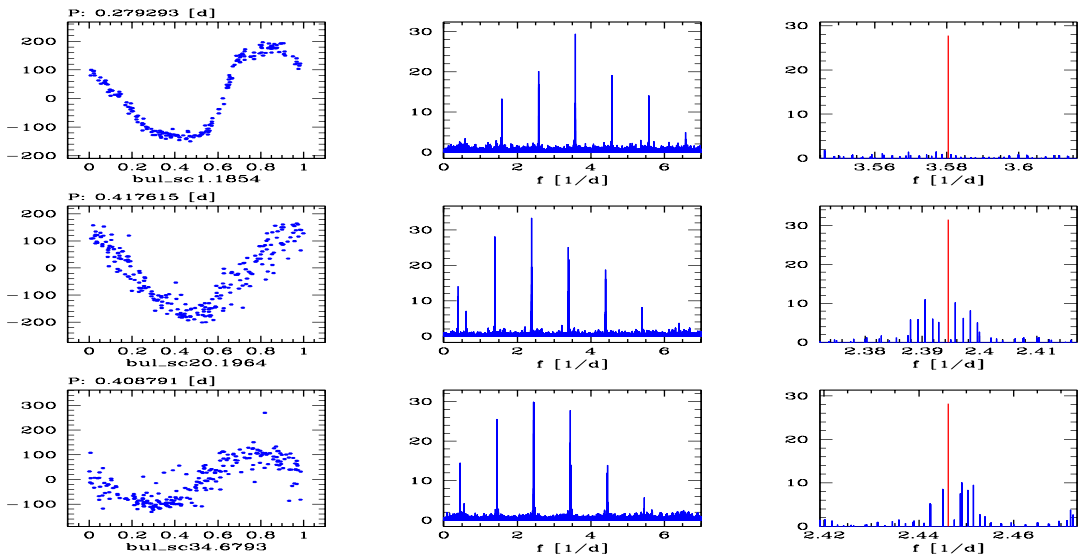}} \par}
\caption{{\small \label{misc}\emph{Stars with multiple additional peaks. Convention is the same as on figure \ref{bla}.}}\small }
\end{figure}

 \hspace{1cm} My sample of multiperiodic RR Lyrae stars contains some objects that can not be classified as either
Blazhko type 1 stars BL1s or type 2 BL2s on the grounds of current
definitions of these classes. Although their light curves look very
similar to light curves of Blazhko stars, with the characteristic scatter
around maximum and minimum of brightness, prewhitened periodograms of these
objects reveal presence of more than two additional peaks. There are also a few objects with frequency triplets, but
unlike in the case of BL2 stars, those triplets are not
equidistant. Existence of RR Lyr stars with such signatures in the
prewhitened spectra was reported by Moskalik \& Poretti (2002), who found a few stars
of this type in OGLE-I database. OGLE-II database contains much more
examples. I found 86 such objects among RRab
stars and 41 among RRc stars. This gives respectively 4.4$\%$ and 5.3$\%$
incidence rates in both classes. Examples are shown in figure \ref{misc}.
\par \hspace{1cm} It is not clear yet if these types of objects could also be referred to
as Blazhko stars. It is possible that in the future they will be called BL3
stars. They certainly share some features with Blazhko stars, but on the
other hand constitute a distinct group. Their nature is uncertain, but they
seem to be another strong evidence for excitation of nonradial modes. 

\subsection{Objects of uncertain nature}

\par \hspace{1cm} The last class of my multiperiodic RR Lyrae catalog
consists of objects that could not be classified reliably as members of any
of the classes described above. Most of these objects are probably BL1 or BL2
stars, but computed modulation periods are longer than the time span of the
data. Additional observations in the future are needed to verify the
hypothesis that these objects are BL1 or BL2 stars with very long
modulation periods. My catalog contains 14 such objects in the RRab class
and 2 objects in the RRc class.  

\section{Summary}

\par \hspace{1cm} I presented the results of an automated search and
analysis of RR Lyrae stars in the Galactic Bulge. My RR Lyrae samples
are very numerous, about 12 times richer than OGLE-I RR Lyrae catalogs, and are therefore well suitable for statistical analysis.
Out of more than 2700 RR Lyrae stars, that I identified in the OGLE-II database,  
about 600 proved to be multiperiodic. Most of the multiperiodic RR Lyrae
stars are Blazhko stars. They show a large variety of pulsation periods and
frequency patterns. The incidence rate of Blazhko stars in the Galactic
Bulge, determined from my analysis, is very high comparing to some other
systems like the LMC and the SMC. With about 25$\%$ in RRab and 10$\%$ in RRc
classes, incidence rates are almost 2.5 times greater than those for the
LMC found by Soszynski et al. (2003). Nature of this phenomenon is unknown and is yet to be understood. High
number of about 100 RRc Blazhko stars is especially interesting because until
now not many variables of this kind were known. They are very different
from RRab Blazhko stars in certain aspects. 
\par \hspace{1cm} On the other hand I have found only
3 RRd stars in addition to the 2 previously known. This means that their incidence rate, with respect to RRc
stars, is $\approx 0.007$, and is about 20 smaller than their incidence rate in the LMC given
by Alcock et al. (2000).
\par \hspace{1cm} As for now we still do not understand why there are so
many more Blazhko stars and so many less RRd stars in the Galactic Bulge
comparing to the Magellanic Clouds. These striking differences call for an explanation.

\section{Acknowledgments}

\par \hspace{1cm} I would like to thank Dr. Wojciech Dziembowski and
Dr. Andrzej Udalski for their helpful comments on
the script and valuable discussions. This work was supported by Polish KBN
grant 5 P03D 030 20.

\end{document}